\documentclass[11pt]{article}

\usepackage[utf8]{inputenc}
\usepackage{amsmath,amsthm,amssymb,enumerate,framed,mdwlist}
\usepackage{color,colortab}
\usepackage[square,numbers,sort&compress]{natbib}

\newcommand{\R}{\mathbb{R}}

\renewcommand{\epsilon}{\varepsilon}

\newtheoremstyle{mythmstyle}
	{\topsep}
	{\topsep}
	{\itshape}
	{}
	{\scshape}
	{.}
	{3pt}
	{}
\theoremstyle{mythmstyle}

\newtheorem{nn}{}[section]

\newtheorem{theorem}[nn]{Theorem}

\newtheorem{REMARK}[nn]{Remark}

\def\ve#1{\mathchoice{\mbox{\boldmath$\displaystyle\bf#1$}}
{\mbox{\boldmath$\textstyle\bf#1$}}
{\mbox{\boldmath$\scriptstyle\bf#1$}}
{\mbox{\boldmath$\scriptscriptstyle\bf#1$}}}

\newcommand{\x}{{\ve x}}

\newcommand{\p}{{\ve p}}
\newcommand{\q}{{\ve q}}

\numberwithin{equation}{section}

\allowdisplaybreaks

\begin{document}

\title{An exposition of special relativity without appeal to ``constancy of speed of light" hypotheses}

\author{Amitabh Basu\footnote{Department of Applied Mathematics and Statistics, The Johns Hopkins University. 
}}
\maketitle

\begin{abstract}
We present the theory of special relativity here through the lens of differential geometry. In particular, we explicitly avoid any reference to hypotheses of the form ``The laws of physics take the same form in all inertial reference frames" and ``The speed of light is constant in all inertial reference frames", or to any other electrodynamic phenomenon. For the author, the clearest understanding of relativity comes about when developing the theory out of just the primitive concept of time (which is also a concept inherent in any standard exposition) and the basic tenets of differential geometry. Perhaps surprisingly, once the theory is framed in this way, one can {\em predict} existence of a ``universal velocity" which stays the same in all ``inertial reference frames". This prediction can be made by performing much more basic time measurement physical experiments that we outline in these notes, rather than experiments of an electrodynamic nature. Thus, had these physical experiments been performed prior to Michelson-Morley type experiments (which, in principle, could have been done in any period with precise enough time keeping instruments), the Michelson-Morley experiments would simply give us an example of a physical entity, i.e., light, which enjoys this special ``universal" status.
\end{abstract}

\section{Introduction}

These notes arose out of the author's attempts to understand the basic principles of relativity (special and general) from the standpoint of a mathematician without any formal training in physics. Due to this handicap, an attempt to understand special relativity from standard physics texts led to confusions and inconsistencies. Most physics textbooks on special relativity proceed on an intuition that most physics undergraduates would have that the author lacked, such as notions like ``inertial frames", ``forces" etc~\cite{bergmann1976introduction,hartle2003gravity,resnickintroduction,rindler-special-relativity,wheeler1966spacetime}. For example, almost all undergraduate (and many graduate) texts on special relativity develop the theory on two principles (stated in slightly different versions in different expositions):

\begin{enumerate} 
\item ({\em Principle of relativity}) All laws of physics take the same form in all inertial reference frames.
\item ({\em Speed of light hypothesis}) The speed of light is the same in all inertial reference frames.
\end{enumerate}

Someone without well-developed physics intuition like the author is immediately led to ask several questions about these hypotheses. 

\paragraph{What is a reference frame?} This is usually not defined very crisply (and most physicists might maintain the position that precise definitions are not even necessary -- any trained physicist has a good understanding of what a reference frame is). The textbooks that do attempt to define a reference frame will make allusions to ``a system of rigid rods and synchronized clocks attached to a material body/observer that can measure the space and time coordinates of events under observation". The most unsatisfactory nature of this definition is that it invokes even more complex concepts like ``rigid rods" and ``synchronized clocks" without much clarification on what these mean. As far as the author can see, what is meant by a reference frame is a coordinate system, i.e., any procedure for assigning 3 ``space" coordinates and 1 ``time" coordinate to events under observation.

\paragraph{What is an inertial reference frame?} The usual answer to this is that an inertial reference frame is one in which any material body that is not under the influence of any forces moves in a straight line~\cite{bergmann1976introduction,resnickintroduction,rindler-special-relativity,wheeler1966spacetime}. First, this already involves another concept of ``force", which is also taken as a primitive, undefined concept. Second, it is not clear how to decide whether a material body is under the influence of any force: one must avoid circularities like ``a material body is free from force when it moves in a straight line in an inertial reference frame". Nevertheless, if one takes the notion of ``force" as a more primitive concept, one can avoid such circularities by positing that one can decide given any material body whether it is under the influence of forces because one understands the source causes of forces. However, a mathematician is also led to the question ``What is a straight line?" One response would be ``A trajectory of a material body is a straight line if the space coordinates in the reference frame are a linear function of the time coordinate". While this seems to pin down a reasonable definition of an ``inertial reference frame" (once one allows for the primitive notion of ``force"), things become tricky when one goes deeper into special relativity. For the reader who knows some basic Newtonian mechanics as well as Special Relativity, the following follow-up question will be interesting: If one performs a {\em Galilean transformation} on the coordinates of an inertial reference frame, one also obtains a coordinate system that satisfies the above definition of an inertial reference frame because Galilean transformations map straight lines to straight lines (as defined above). However, in relativity, only the Lorentz transformations are claimed to map inertial reference frames to inertial reference frames. Not Galilean transformations. One may argue that indeed, in the definition of the inertial reference frame, one has to impose the condition that light has the same velocity on all inertial reference frames. In the author's opinion, such a defense comes dangerously close to circularity: The hypothesis 2. above that forms the foundation for deriving special relativity seems to be now part of the {\em definition} of an inertial reference frame rather than a {\em physical hypothesis}. I am not sure if this viewpoint will be endorsed by any physicist.

There are other definitions in many textbooks (e.g., which appeal to ``Cartesian systems of coordinates"~\cite{bergmann1976introduction,hartle2003gravity}) which, for the author, are also not free from similar imprecision (e.g., what exactly makes a coordinate system ``Cartesian" is not always clear).

\paragraph{What is meant by ``laws of physics take the same form in all inertial reference frames"?} This is almost never explained clearly in any textbook that introduces special relativity and was the biggest source of the author's confusion in initial attempts to understand relativity. In fact, the author has not been able to find any physics textbook with a clear exposition of what this phrase means either as a physical hypothesis or a mathematical assumption/definition. First, it is already difficult to pin down what is meant by a ``law of physics". Second, layered notions like ``take the same (mathematical?) form" are notoriously tricky to make precise even within mathematics. 

 \bigskip
 \bigskip

The author would like to clarify that these questions are an impediment to understanding relativity only for an outsider like himself. Physicists have a perfectly clear grasp of relativity without having to bother about precise definitions and meanings of concepts that they have an instinctive understanding of, owing to their extensive training in physical experiments (real or thought) and discussions with their fellow scientists. For someone who has not been exposed to this training and community wisdom, and therefore lacks the physicist's intuition, it then becomes hard to get to the essence of relativity theory on the basis of these ideas and the standard expositions quickly become impenetrable.

The author's route to understanding relativity was based on the concepts of differential geometry, a topic which is is much closer to the author's educational training. The only primitive concept that is used is that of ``time". No other epistemologically complex notions such as ``inertial frames", ``invariance of laws of physics", ``speed of light", ``force" and so on are needed. Of course, reliance on these concepts is replaced by the crutch of (very basic) differential geometry. These notes are an attempt to commit this approach to paper. 

We do not claim novelty of this approach. The textbooks of Synge~\cite{synge1956relativity,synge1960relativity} essentially take the same approach with ``time" and differential geometry taking center stage in relativity. The same can be said about Eddington's original exposition, written soon after Einstein published his results~\cite{eddington1923the}. For example, the notion of an ``inertial reference frame" is not even mentioned once in Synge's textbook on special relativity~\cite{synge1956relativity}, and only mentioned tangentially 3-4 times in Eddington's book. Ironically, to the author, the clearest definition of an ``inertial reference frame" falls out naturally from this approach, and thus the notion can be simply based on the notion of ``time" and basic differential geometry ideas. Another source which closely follows this approach are the lecture notes of Robert Geroch~\cite{geroch2013general}. It is interesting for the author that this route to explaining relativity is found in the minority of textbooks and expositions on the subject. The author clearly has his own bias, but it would make an interesting debate as to whether the more physical notions of ``inertial frames", ``forces", ``invariance of laws" are superior or inferior to differential geometry as a foundation on which to erect the special theory of relativity. Of course, once we move to the general theory, since differential geometry is indispensable, one might as well begin there.

A lesser publicized aspect of this approach is that, in hindsight, {\em certain simple time measurement experiments can be designed that would \underline{predict} the existence of a ``universal velocity" that does not change when we change ``inertial reference frames" (once an appropriate and precise definition of ``inertial frames" is obtained);} see Section~\ref{sec:universal-speed} below. Thus, if these experiments had been performed prior to the Michelson-Morley experiments, an alternative route to special relativity would most likely have been found and the Michelson-Morley experiments would have simply exhibited a concrete physical entity that possesses this universal velocity and provided further confirmation of the theory, rather than being the starting point of the development of the theory. We quote Eddington: ``It is {\em shown} that there must be a particular speed which has the remarkable property that its value is the same for all these [inertial] systems; and by appeal to the Michelson-Morley experiment or to Fizeau's experiment is found that this is a distinctive property of light ..." \cite[pp. 41--42]{eddington1923the} (the emphasis and the parenthetical word is ours). 

The only difference in these notes and the prior works of Synge, Eddington and Geroch cited above is that we make use of Sylvester's ``Law of Inertia" from Linear Algebra, which is absent from all these manuscripts. It is the author's opinion that the use of this basic linear algebraic fact makes the development of the theory even more natural and clear.

\section{Events and (Proper) Time}

The most fundamental concept in relativity is that of an {\em event}. These are idealizations of real physical occurrences in the limit that the duration of the phenomenon is infinitesimal. The goal of relativity is to give quantitative relationships between events. This is done by first assigning {\em coordinates} to the set of all possible events. The notion of ``all possible events" is decidedly vague; however, it is no more vague than the concept of ``space" which corresponds to the ``set of all possible locations". Thus, the main hypothesis of relativity is that physics takes place on the canvas of ``all possible events". Some of the elements of this set correspond to ``real" physical occurrences or events, and the rest are hypothesized elements which only act as ``glue" between the actual physical events. In any case, from a mathematical point of view, the basic notion is that of a set, the elements of which are to be called {\em events}. The question now arises as to how to assign coordinates to events, and more primitively, how many coordinates are to be assigned? We will come back to this once we discuss our second primitive concept: time.

Certain subsets of events in the ``real world" are seen to be connected in an intimate way to each other and together they give the history of a physical entity, like an electron or a neutron. The events associated in such a way is called a {\em worldline}. Now, it is a matter of experience that any entity associated with a worldline has an ability to ``keep time". Herein lies the second main assumption in relativity: there exist certain cyclic processes in this world that can be used to assign real number ``lengths" or ``time intervals" between any two events on a given worldline: the number of cycles that have repeated between these two events. Moreover, there is an ordering of the events on this worldline. Thus, one can use the real numbers to label all events on the worldline. More precisely, one can\footnote{In these notes, whenever we use phrases like ``one can", this is an idealization of the process of setting up physical experiments or apparatus.} set up a one-to-one correspondence between events on a wordline and the real numbers, such that 1) the ``time interval" between any two events is the absolute difference of the real numbers labeling these two events, 2) an event with a smaller label is said to ``occur before" an event with a larger label, and 3) an arbitrary event on the worldline is given the label 0. The standard example of such a time keeping device is that of a Cesium clock. 
It is important to note here that the real numbers are assigned to events that happen {\em on the worldline of the clock}.  

This is the point where we can start to make an appeal to differential geometry ideas and come back to the question of assigning {\em coordinates} to events. We want to impose the structure of a differentiable manifold on the set of events, because this mathematical machinery allows one to assign ``lengths" to ``curves" in a set. For this, we require mechanisms to assign ``coordinates" to events. A concrete procedure is the following. Consider the worldline and corresponding events associated with a Cesium clock and think of the clock as an ``observer". Consider any event $E$ that is not on the worldline of the observer. We posit that any such $E$ can be ``observed" at the Cesium clock by means of some signals, e.g., a particle moving from $E$ to an event on the worldline [more precisely, a particle whose worldline goes through $E$ and intersects the worldline of the observer], or a sound wave or light wave starts at $E$ and ends at an event on the worldline. Then the observer can assign the same real number to event $E$ as the number it sees on the clock when the ``signal" from $E$ reachers the clock. By this means, the ``observer" can assign numbers to events that can possibly be connected to it in the chosen way (which may be a strict subset of the entire set of possible events). Thus, if we have $n$ Cesium clocks along with their own individual worldlines, we have $n$ observers, and hence $n$ coordinates on the set of events (possibly a strict subset of events). One can have many other ways of assigning coordinates to events. For instance, in classical mechanics, one assumes ways of assigning three ``space" coordinates using standardized scales and a ``time" coordinate to each event. We allow any such mechanism to assign coordinates to events; the only criterion (for now) is that each distinct event is assigned a unique value by a coordinate assignment procedure (we will impose a second criterion of ``smoothness" below, which is harder to formalize in physical terms).

The central hypothesis of relativity theory is the following:

\begin{quote} {\bf The set of events is a 4-dimensional differentiable manifold. The time interval between two events on any worldline in this manifold is determined by a pseudo-metric on the manifold. This manifold is called {\em spacetime}.}
\end{quote}

Thus, we assume that {\em four} independently moving Cesium clocks are enough to assign coordinates to the set of events; the above hypothesis implies that the coordinates assigned by any other Cesium clock can be obtained as a function of the $4$ coordinates already assigned. A permissible coordinate assigning procedure needs to be ``smooth" with respect to the ``clock mechanism" of assigning coordinates (mathematically, we are talking about the function having derivatives of all orders). From the set of all permissible coordinate assigning mechanisms, one can consider any 4 independent ones (i.e., no one mechanism is a function of the other three) to assign coordinates to the set of events.

\section{The pseudo metric}

Given that we have 4 independent coordinate assigning mechanisms in place, one can now physically probe the properties of the hypothesized pseudo metric that measures time intervals. This is modeled in differential geometry in the following way: At each event or point $E$ of the manifold, one has a symmetric bilinear function $G(E)$ defined on the tangent vector field at $E$ that varies ``smoothly" from one event to another (more precisely, one has a smooth second order tensor field on the manifold). The ``distance" between two nearby points $v$ and $v + \Delta v$ in the manifold is approximately given by $G(v)(\Delta v, \Delta v)$; the approximation gets better and better as $\Delta v \to 0$ (the author is deliberately avoiding precise mathematical definitions here to get the idea across). Given an arbitrary coordinate system for the manifold, the function $G(v)$ at any point $v$ with coordinates $\x = (x_1, x_2, x_3, x_4)$ is given by a symmetric $4\times 4$ matrix $g(\x)$ with entries $g_{ij}(\x)$, $i, j = 1, \ldots, 4$, and the distance between $\x$ and $\x + \Delta \x = (x_1 + \delta x_1, x_2 + \delta x_2, x_3+\delta x_3, x_4 + \delta x_4)$ is given by \begin{equation}\label{eq:metric-coordinate}\sum_{i,j=1}^4 g_{ij}\delta x_i \delta x_j.\end{equation} More generally, given any curve $C(\gamma)$, $a \leq \gamma \leq b$ in spacetime, one defines the {\em squared length} of the curve as \begin{equation}\label{eq:length-curve}s\ell(C):= \int_a^b G(C(\gamma))(C'(\gamma), C'(\gamma)) d\gamma.\end{equation}
Note that the squared length of a curve, as defined above, can be negative because there is no guarantee on what the sign of $G(\gamma(t))(\gamma'(t), \gamma'(t))$ is, in general. Thus, the nomenclature ``squared length" is perhaps not ideal, but we will use this because when the curve is a worldline, the squared length is positive and gives the square of the time interval between then events $C(a)$ and $C(b)$.

Different coordinate systems will give different values for the entries of $G(v)$ (just like different coordinate systems will give different values for the coordinates of $v$). Given a $4\times 4$ symmetric matrix, one can compute the (real) eigenvalues of this matrix. Some of these eigenvalues will be positive, some negative and some zero; we will denote the number of such eigenvalues by $n_+, n_-, n_0$ respectively. The tuple $(n_+, n_-, n_0)$ is called the {\em signature} of the matrix.

A basic result in linear algebra/differential geometry is that no matter what coordinate system is used, the signature of the matrix $g(v)$ never changes (even though the entries of the matrix certainly change; in fact the coordinate of $v$ also change). Here is the main observation:

\begin{quote} {\bf Once we have made the basic assumption of modeling events as a 4-dimensional differentiable manifold with a pseudo-metric, one can use physical experiments to determine the signature of the pseudo metric at any event. One way to do this is the following. Since the signature is invariant with respect to coordinate systems, one first determines the entries $g_{ij}(v)$ of the pseudo metric matrix at the event $v$ using physical experiments, and then computes the eigenvalues and thus the signature. Which physical experiments? Since the pseudo metric, and thus the corresponding matrix, is assumed to be symmetric one needs to determine only $10$ of the upper triangular entries.  One may determine these entries by ``shooting off" $10$ independent Cesium clocks from $v$ (thus, all 10 worldlines (approximately) pass through $v$), and recording the coordinates of the events on these Cesium clocks a tiny time interval later (as per the observer's own Cesium clock, say). We know the time intervals recorded by the Cesium clocks at these $10$ future events. One plugs into~\eqref{eq:metric-coordinate} the $\delta x_i$'s for these $10$ events and equates them to the corresponding time intervals recorded. We now have 10 independent equations in the $10$ entries that we want to solve for. This gives us the metric.}
\end{quote}

If the above idealized experiment were to be performed, one would observe that the signature of the pseudometric everywhere is $(1,3,0)$, i.e., one positive eigenvalue, 3 negative eigenvalues and no zero eigenvalues. Such a pseudo metric is called the {\em Minkowski metric}.

\section{Inertial reference frames, observer's space and time}

Another basic fact of differential geometry/linear algebra is that given any event $E$, one can always choose coordinate systems for the manifold such that the pseudo metric matrix $g(E)$ at $E$ becomes a diagonal matrix with $\pm 1$ on the diagonal. In particular, with signature $(1,3,0)$, there exists a coordinate system (that depends on the event $v$) such that $g_{ij}(v) = 0$ if $i\neq j$, $g_{11} = 1$, and $g_{22} = g_{33} = g_{44} = -1$.

We define such a coordinate system as an {\em inertial coordinate system associated with $v$}. Note that, in general, in an inertial coordinate system associated with $v$, the pseudo metric at other events may have a matrix that is {\em not diagonal}. Characterizing when this does or does not happen needs the concept of {\em curvature} of a differentiable manifold with a pseudo metric. In fact, a fundamental theorem is the following~\cite{boothby1986introduction}.

\begin{theorem}\label{thm:flat-coordinates} 
Let $M$ be a differentiable manifold with a pseudo metric. Then there exists a coordinate system which makes the pseudo metric diagonal (with $\pm 1$ on the diagonal) everywhere if and only if the curvature is zero everywhere.
\end{theorem}

In the above, we have not formally defined curvature; but this is not essential for the discussion in these notes. What is important is the following.

\begin{quote} The difference between special relativity and general relativity is that one assumes in special relativity that spacetime is {\em flat}, i.e., zero curvature everywhere, whereas in general relativity, this assumption is dropped. Equivalently, by Theorem~\ref{thm:flat-coordinates}, in special relativity, one assumes the existence of a coordinate system for all of spacetime such that the pseudo metric matrix is diagonal (with $\pm 1$ on the diagonal) at every event.
\end{quote}

We remain in the realm of special relativity in the remainder of these notes. Thus, we can speak of {\em inertial reference frames} without the need to mention a particular event $v$. To summarize:

\begin{quote} {\bf An inertial reference frame is a coordinate system on spacetime such that the pseudo metric takes the canonical diagonal form everywhere.}
\end{quote}

It is not hard to see that an inertial reference frame is not unique. In fact, there are infinitely many different inertial reference frames. We tackle this infinity of choices in the next section. Before that, we state a very interesting characterization of inertial reference frames that, at least to the author, connects the abstract mathematical definition above to the more common physical intuition of ``inertial reference frames" from physics textbooks.

A {\em timelike curve} is a curve in spacetime such that the tangent vector at any point on the curve has positive length as per the pseudo metric. A {\em geodesic} is a timelike curve $C$ such that for any two points $p,q$ on the curve $C$, the length of the curve between $p$ and $q$ (as computed by~\eqref{eq:length-curve}) reaches a {\em critical value} (e.g. maximum or minimum) amongst all curves passing through $p$ and $q$. The following theorems can be derived from standard results in differential geometry~\cite{boothby1986introduction}.

\begin{theorem}\label{thm:geodesic-inertial-1} Let $S$ be a flat spacetime, i.e., there exist inertial coordinate systems. 

Let $(x_1, x_2, x_3, x_4)$ be an inertial coordinate system. The curve given by setting $x_2 = x_3 = x_4 = 0$ is a geodesic in spacetime. 

Conversely, let $C(\gamma)$, $\gamma\in \R$ be a geodesic in $S$. Then there exists an inertial reference frame $(x_1, x_2, x_3, x_4)$ such that $C$ corresponds to the set of points $x_2 = x_3 = x_4 = 0$.
\end{theorem}

\begin{theorem}\label{thm:geodesic-inertial-2} Let $S$ be a a flat spacetime and let $(x_1, \ldots, x_4)$ be an inertial reference frame. Any curve $C(\gamma)$ is a geodesic if and only if it is given by affine linear functions of $\gamma$, i.e., there exists a vector $d \in \R^4$ and a point $v \in S$ such that in the coordinates $(x_1, \ldots, x_4)$, $C(\gamma) = v + \gamma d$ for all $\gamma \in \R$.
\end{theorem}

By Theorem~\ref{thm:geodesic-inertial-1}, one can associate a Cesium clock/observer with the $x_1$ axis of an inertial reference frame. We will refer to this as the {\em observer associated with an inertial reference system}. We now make several interesting observations.

\begin{enumerate} 
\item Given two events $\p$ and $\q$, there is a unique geodesic $C(\gamma)$ with endpoints $p$ and $q$; by Theorem~\ref{thm:geodesic-inertial-2} the coordinates of the curve $y_i(\gamma)$ are linear functions of $\gamma$ in this inertial system. We say the {\em squared distance} between $p$ and $q$ is the squared length of $C$, i.e., $s\ell(C)$. If $p$ and $q$ have coordinates $(p_1, p_2, p_3, p_4)$ and $(q_1, q_2, q_3, q_4)$,~\eqref{eq:length-curve} gives the squared distance for this geodesic as $$(p_1 - q_1)^2 - (p_2 - q_2)^2 - (p_3 - q_3)^2 - (p_4 - q_4)^2.$$

\item For any two events on the $x_1$ axis of this inertial reference frame, the time interval as measured by the clock, is precisely the difference between the $x_1$ coordinates of these events (see 1. above). We will make the natural extension to say that any event with coordinates $(y_1, y_2, y_3, y_4)$ in this inertial system {\em occurred at time $y_1$}. 

\item Consider two events that occurred at the same time with coordinates $(t, w, y, z)$ and $(t, w', y', z')$. By the above discussion, the squared distance between these two events is $-[(w-w')^2 + (y-y')^2 + (z - z')^2)]$ (see 1. above). This coincides (modulo a negative sign) with the standard three dimensional Euclidean metric. Thus, we identify all events in this inertial system with the same time coordinate to be a 3 dimensional {\em space} associated with that particular time instant. One could extend this to define a {\em spatial distance} between two events with {\em different} time coordinates, with respect to this inertial reference frame: given events $(t, w, y, z)$ and $(t', w', y', z')$, the spatial distance between these events is $[(w-w')^2 + (y-y')^2 + (z - z')^2)]^{1/2}$.
\end{enumerate}


Theorem~\ref{thm:geodesic-inertial-1} and the enumerated comments above show how {\em space} is a derived concept in relativity from the more fundamental notion of {\em time}. Also, Theorem~\ref{thm:geodesic-inertial-2} connects the mathematical definition of inertial frames in these notes to the more commonly used notion of ``coordinate system associated with material bodies moving in a straight line" in physics texts.

\section{A universal speed}\label{sec:universal-speed}

We now derive a physically non-intuitive and surprising fact out of the above concepts. Consider two inertial coordinate systems $(x_1, \ldots, x_4)$ and $(x'_1, \ldots, x'_4)$. Consider a geodesic in the first system by $C(\gamma)$ such that for any two points on this line, the squared distance between these two points is $0$. For instance the curve $C(\gamma) = (\gamma, \gamma, 0, 0)$ for $\gamma \in \R$ satisfies this property. Now imagine an entity traversing this geodesic in spacetime. One can define its ``speed" by considering any two points and dividing the spatial distance (as described by point 3. above Theorem~\ref{thm:geodesic-inertial-2})  between these two points by the difference in time coordinates of these two events. However, since the squared distance is $0$, it is not hard to see that this ratio is $1$ in this inertial reference frame. However, when we change inertial reference frames, the squared distance, of course, does not change: this is a property of the curve and the pseudo metric of spacetime, {\em not the coordinates}. Consequently, the speed in the other inertial frame is ALSO $1$! Thus, we come to the existence of a {\em universal speed} associated with the spacetime of special relativity (i.e., pseudo metrics of signature $(1, 3,0)$), which is invariant in different inertial reference frames.

More generally, through any point $\p$ in spacetime with coordinates $(p_1, p_2, p_3, p_4)$ in some inertial reference frame, one can define a cone of points $(x_1, x_2, x_3, x_4)$ satisfying $(p_1 - x_1)^2 - (p_2 - x_2)^2 - (p_3-x_3)^2 -(p_4 - x_4)^2 = 0$. This is the union of all geodesics passing through $\p$ satisfying the above property that the squared distance between any two points on the geodesic is $0$. Any entity traversing such a geodesic will have the same speed in every inertial reference frame. This is refereed to as the {\em light cone} in relativity literature.

\section{Lorentz transformations}\label{sec:lorentz} We now deal with the question of non unique inertial reference frames. Using standard results in linear algebra, one can show that any two inertial coordinate systems $(x_1, \ldots, x_4)$ and $(x'_1, \ldots, x'_4)$ are related by an affine transformation. In particular, there exists a matrix $L$ with entries $L_{ij}$, $i,j,=1, \ldots, 4$ and $t \in \R^4$ such that if for any event $E$ with coordinates $\x$ and $\x'$ in the two systems, we have $\x' = L\x + t$. Any such transformation is called a {\em Lorentz transformation}. We will derive a special form of the Lorentz transformation to illustrate the point. 

Consider two inertial reference systems $(x_1, \ldots, x_4)$ and $(x'_1, \ldots, x'_4)$. By Theorem~\ref{thm:geodesic-inertial-1}, the $x'_1$-axis is a geodesic in spacetime. By the Theorem~\ref{thm:geodesic-inertial-2}, the $x'_1$-axis is a straight line in the coordinate system $(x_1, x_2, x_3, x_4)$. The coordinate system $(x_1, x_2, x_3, x_4)$ imposes the $\R^4$ vector space structure on spacetime. One can then consider the two dimensional subspace $V$ spanned by the $x_1$-axis and the line corresponding to the $x'_1$-axis. This subspace $V$ intersects the 3-dimensional ``spatial" subspaces of the two inertial frames, i.e., points satisyfing $x_1 = 0$, and points satisfying $x'_1 = 0$, respectively in two lines $\ell$ and $\ell'$. Since the spatial metrics are the standard 3 dimensional metrics, one can always perform a rotation in the respective subspaces such that the new $x_2$ axis in the $(x_1, x_2, x_3, x_3)$ system coincides with $\ell$ and, similarly, the new $x'_2$ axis in the $(x'_1, x'_2, x'_3, x'_3)$ system coincides with $\ell'$ line. Thus, we now have a two dimensional subspace $V$ of spacetime, which contains the $x_1, x_2$ axes, as well as the $x'_1, x'_2$ axes. Moreover, one can perform additional rotations in the $x'_3, x'_4$ space such that these axes coincide with the $x_3$ and $x_4$ axis (since the spatial metric is the standard Euclidean metric). We now have that for any point $\p$ in spacetime, the coordinates $(p_1, p_2, p_3, p_4)$ and $(p'_1, p'_2, p'_3, p'_4)$ satisfy $p_3 = p'_3$ and $p_4 = p'_4$. Additionally, we must have \begin{equation}\label{eq:hyperbolic}(p_1)^2 - (p_2)^2 = (p'_1)^2 - (p'_2)^2.\end{equation}
Now consider the $x'_1$-axis which lies in the $x_1, x_2$ plane in the unprimed coordinates. In the unprimed coordinates, let the equation of this straight line be $x_2 = vx_1$ for some real number $v$. Consider any point with coordinate $(1,0,0,0)$ in the primed coordinates on this straight line, with corresponding coordinates $(p_1, p_2,0,0)$ in the unprimed coordinates. By~\eqref{eq:hyperbolic}, we obtain $$1 = p_1^2 - p_2^2 = p_1^2(1- v^2),$$ which implies that the coordinates of $(1,0,0,0)$ in the unprimed coordinates are $(\frac{1}{\sqrt{1-v^2}}, \frac{v}{\sqrt{1-v^2}}, 0,0)$. Note that this imposes the condition that $|v| < 1$ on the $x'_1$-axis equation. This corresponds to the condition that the speed of any entity that can define the time axis of an inertial frame must be less than $1$, which is the universal speed from Section~\ref{sec:universal-speed}. 

Let the $x'_2$-axis be defined by $\bar{v}x_2 = x_1$. By a similar calculation as above, the point $(0,1,0,0)$ in the primed coordinate has coordinates $( \frac{\bar{v}}{\sqrt{1-\bar{v}^2}},\frac{1}{\sqrt{1-\bar{v}^2}}, 0,0)$. We thus have arrived at our coordinate transformations:

\begin{equation}\label{lorentz-prelim}
\begin{array}{l}
x_1 = \frac{x'_1}{\sqrt{1-v^2}} + \frac{\bar v x'_2}{\sqrt{1-\bar{v}^2}} \\
x_2 = \frac{v x'_1}{\sqrt{1-v^2}} + \frac{x'_2}{\sqrt{1-\bar{v}^2}} \\
x_3 = x'_3 \\
x_4 = x'_4
\end{array}
\end{equation}

Since the pseduometric is preserved in both coordinate systems, the {\em inner product} with respect to this pseudometric is also preserved. In particular, the inner product between $(1,0,0,0)$ and $(0,1,0,0)$ in the primed coordinates is $0$. Thus, the inner product in the unprimed coordinates must also equal $0$. The inner product is $\frac{\bar v}{\sqrt{1-v^2}\sqrt{1-\bar{v}^2}} - \frac{v}{\sqrt{1-v^2}\sqrt{1-\bar{v}^2}}$. Equating this to $0$ implies that $v = \bar v$. Therefore, from~\eqref{lorentz-prelim}, we obtain the transformations:

\begin{equation}\label{lorentz}
\begin{array}{l}
x_1 = \frac{x'_1 +v x'_2}{\sqrt{1-v^2}} \\
x_2 = \frac{v x'_1+x'_2}{\sqrt{1-v^2}} \\
x_3 = x'_3 \\
x_4 = x'_4
\end{array}
\end{equation}
The inverse transformations are computed by taking the matrix inverse:

\begin{equation}\label{eq:canonical-lorentz-matrix}
\hat L:= \left[ \begin{array}{cccc} \frac{1}{\sqrt{1-v^2}} & \frac{v}{\sqrt{1-v^2}} &0&0 \\
\frac{v}{\sqrt{1-v^2}}& \frac{1}{\sqrt{1-v^2}} & 0 & 0 \\
0 & 0 & 1 & 0 \\
0 &0 &0&1
\end{array}\right]^{-1} = \left[ \begin{array}{cccc} \frac{1}{\sqrt{1-v^2}} & \frac{-v}{\sqrt{1-v^2}} &0&0 \\
\frac{-v}{\sqrt{1-v^2}}& \frac{1}{\sqrt{1-v^2}} & 0 & 0 \\
0 & 0 & 1 & 0 \\
0 &0 &0&1
\end{array}\right]
\end{equation}
[Note that this corresponds to simply setting $v$ to $-v$, capturing the fact that if the primed system is moving with speed $v$ with respect to the unprimed system, then the unprimed system is moving with speed $-v$ with respect to the primed system.]

\noindent Thus, the inverse Lorentz transformations become 

\begin{equation}\label{lorentz}
\begin{array}{l}
x'_1 = \frac{x_1 -v x_2}{\sqrt{1-v^2}} \\
x'_2 = \frac{x_2 - vx_1}{\sqrt{1-v^2}} \\
x'_3 = x_3 \\
x'_4 = x_4
\end{array}
\end{equation}
Or in the more familiar form

\begin{equation}\label{lorentz}
\begin{array}{l}
t' = \frac{t -v x}{\sqrt{1-v^2}} \\
x' = \frac{x - vt}{\sqrt{1-v^2}} \\
y' = y \\
z' = z
\end{array}
\end{equation}

Since we also had rotations involved in obtaining the above canonical transformations, a general Lorentz transformation is given by $\x' = L\x + t$ where $t$ is an arbitrary vector in $\R^4$ and $L = \hat O_1 \hat L \hat O_2$, where $\hat L$ is from~\eqref{eq:canonical-lorentz-matrix} and $\hat O_i$ is a matrix of the form $$\left[\begin{array}{cc}1 & \mathbf{0}  \\ \mathbf{0} & O_i\end{array}\right],$$ where $O_i$, $i=1,2$ is a $3\times 3$ orthogonal (rotation) matrix.

\section{Miscellaneous Remarks}
All formulas and derivations in special relativity are based on the Lorentz transformation. Thus, once these have been obtained using our approach as done in Section~\ref{sec:lorentz}, the rest of the theory follows along standard lines, including kinematics (time dilation, length contraction, additional of velocities, twin paradox etc.) and dynamics once the notion of energy/mass has been introduced. We do not deal with the development of dynamics in these notes.



\bibliographystyle{plain}
\bibliography{../../references/full-bib}

\end{document}